\documentclass[journal]{IEEEtran}

\usepackage{amsmath}
\usepackage{amsfonts}
\usepackage{amssymb}
\usepackage{graphicx}
\usepackage[hyphens]{url}     
\usepackage[hidelinks]{hyperref} 
\hypersetup{breaklinks=true}  
\usepackage{color}            

\begin{document}

\title{Pingmark: A Textual Protocol for Universal Spatial Mentions}

\author{
\IEEEauthorblockN{Kalin Dimitrov}
\IEEEauthorblockA{Independent Researcher\\
Veliko Tarnovo, Bulgaria\\
Email: \textit{s2509014279@sd.uni-vt.bg}}
}

\maketitle
\markboth{Preprint --- Pingmark Protocol Specification (PPS v0.1)}{K.~Dimitrov}

\begin{abstract}
Pingmark defines a universal textual protocol for expressing spatial context through a minimal symbol: \textbf{!@}.
Rather than embedding coordinates or using proprietary map links, Pingmark introduces a semantic trigger that compliant
client applications interpret to generate a standardized resolver link of the form
\url{https://pingmark.me/}\texttt{<latitude>/<longitude>[/<timestamp>]}.
This allows location expression to function like existing textual conventions --- \texttt{@} for identity or
\texttt{\#} for topics --- but for physical space. The protocol requires no user registration, relies on open mapping
technologies, and protects privacy by generating location data \textbf{ephemerally and locally}.
This paper presents the motivation, syntax, and design of the Pingmark Protocol Specification (PPS v0.1),
its reference resolver implementation, and the long-term goal of establishing Pingmark as an open Internet
standard for spatial mentions.
\end{abstract}


\section{Introduction}
Digital communication today treats location as a platform feature, not a language element. Every major messenger provides its own ``Share Location'' function, often buried behind multiple steps and incompatible across ecosystems. As a result, expressing position within text remains fragmented and requires action rather than simple syntax. Pingmark reimagines location as a semantic token, not an action. Typing the shorthand \textbf{!@} conveys the intent ``I am at'' in a form any compliant system can interpret. A browser extension, keyboard add-on, or chat parser can automatically generate a corresponding resolver link showing the sender's current position on an open map. The same text remains meaningful even outside these systems, preserving human readability. This work introduces the Pingmark Protocol Specification v0.1, which defines the syntax, resolution rules, and implementation layers of the protocol. It also outlines how Pingmark differs from conventional map-link services by emphasizing textual semantics, openness, and client-side, privacy-preserving data generation.

\section{Related Work}
Several technologies address geolocation exchange, yet none function as open textual syntax allowing client-side, real-time generation:
\begin{itemize}
    \item \textbf{geo: URI (RFC 5870)} --- a formal standard for coordinate expression, but it lacks an intuitive conversational syntax and requires manual entry or complex parsing, making it unsuitable for real-time text input.
    \item \textbf{Google Maps / Apple Maps links} --- closed formats tied to specific apps and data ecosystems.
    \item \textbf{Plus Codes (Open Location Code) and what3words} --- static addressing systems that identify a fixed point, rather than serving as a real-time, ephemeral location mention.
    \item \textbf{@ and \# symbols} --- successful precedents for universal semantic tokens across platforms, proving the viability of the minimal textual trigger approach.
    \item \textbf{W3C Open Annotation Model} --- demonstrates how textual markers can be formalized as interoperable web semantics, motivating lightweight, syntax-level standards.
\end{itemize}
Pingmark extends this linguistic lineage by defining a token for space --- a lightweight, human-friendly standard enabling open interpretation of location intent within any text medium.

\section{Methodology / Design --- Pingmark Protocol Specification (PPS v0.1)}
\textbf{Purpose}---The Pingmark Protocol provides a single, minimal syntax \textbf{!@} that represents a spatial mention inside text. When detected, compliant clients automatically generate a standardized resolver link pointing to an open map visualization. The mechanism separates the \textbf{textual intent} (\textbf{!@}) from the \textbf{data generation} (coordinates).

\textbf{Syntax}---\textbf{!@} acts as a trigger meaning ``I am at.'' It carries no embedded coordinates; generation of location data (latitude/longitude) occurs automatically and locally on the client side when the symbol is typed or detected.

\textbf{Automatic Resolution Rule}---Detected \textbf{!@} $\to$
\url{https://pingmark.me/<latitude>/<longitude>[/<timestamp>]}.

\textbf{Parameters}---\textbf{latitude}, \textbf{longitude}, and optional \textbf{timestamp} (ISO 8601) are handled internally by the client application and are never manually typed by the user. The timestamp is crucial for distinguishing real-time (ephemeral) mentions from past locations.

\textbf{Privacy and Ephemerality}---No user identifiers or accounts are required. The link is generated locally using the device's current location; coordinates are never stored centrally. Resolvers may implement short-term caching but must not retain personal data.

\textbf{Protocol Layers}---Pingmark Protocol (PPS v0.1) $\to$ Symbol \textbf{!@} $\to$ Client (Generates coordinates) $\to$ Resolver (\texttt{pingmark.me}).

\textbf{Implementation Levels}---L1 Parser (browser/keyboard extension that detects \textbf{!@}), L2 Resolver (open map that displays the point), L3 SDK/API (integration into existing platforms).

\subsection{Potential Use Cases}
The design of Pingmark allows seamless integration into diverse communication environments:
\begin{itemize}
    \item \textbf{Instant Messaging:} A user quickly shares their live meeting spot: ``We're waiting at the south entrance \textbf{!@}.'' The client replaces \textbf{!@} with the live, ephemeral link.
    \item \textbf{Email/Documents:} A team lead inserts a specific worksite location into a project update without relying on proprietary maps.
    \item \textbf{Micro-blogging/Social Media:} Users can tag their current event location without the platform needing native location services, relying solely on the textual syntax.
\end{itemize}

\section{Reference Implementation (pingmark.me)}
The resolver \texttt{pingmark.me} serves as a reference implementation of the PPS. It visualizes any compliant link using open-source mapping (OpenStreetMap + Leaflet.js). When opened, the page displays the specified point, optional timestamp, and quick-access buttons such as ``Open in Maps'' or ``Get Directions.'' No cookies, accounts, or analytics are required. This implementation demonstrates the feasibility of the protocol without creating dependency on a single domain.

\section{Discussion and Evaluation}
Initial informal testing showed that users intuitively interpret \textbf{!@} as ``I am at.'' The simplicity of a two-character symbol makes adoption frictionless. Because parsing and coordinate generation occur client-side, Pingmark can integrate into browsers, chat apps, or text editors without compromising user privacy policies.

\textbf{Advantages}---Universal textual syntax; open resolver model; privacy-preserving (ephemeral and local data); backward-compatible even if unparsed.

\textbf{Limitations}---Requires widespread parser support; potential resolver fragmentation without unified governance. Addressing this requires moving towards formal standardization.

\section{Conclusion and Future Work}
Pingmark establishes a minimal yet expressive protocol for spatial mentions, turning location into a first-class textual concept. By separating syntax (\textbf{!@}) from implementation (the resolver), it achieves universality similar to @ and \# for identity and topics. \textbf{Future Work}---Publishing a formal, RFC-style specification (PPS v1.0) and pursuing governance through a standards body like the IETF or W3C to ensure the protocol remains open and unfractured; defining a dedicated \texttt{pingmark://} URI scheme; developing SDKs and plugins for messaging and IoT.

\end{document}